\documentclass[manuscript]{acmart}
\AtBeginDocument{%
  }

\copyrightyear{2025}
\acmYear{2025}
\setcopyright{rightsretained}
\acmConference[OZCHI '25]{37th Australian Conference on Human-Computer Interaction (HCI)}{November 29-December 03, 2025}{Sydney, Australia}
\acmBooktitle{37th Australian Conference on Human-Computer Interaction (HCI) (OZCHI '25), November 29-December 03, 2025, Sydney, Australia}
\acmPrice{}
\acmDOI{10.1145/3764687.3767275}
\acmISBN{979-8-4007-2016-1/25/11}

\begin{document}

\title{Advancing Interdisciplinary Approaches to Online Safety Research}

\author{Senuri Wijenayake}
\affiliation{%
  \institution{RMIT University}
  \city{Melbourne}
  \country{Australia}}
\email{senuri.wijenayake@rmit.edu.au}

\author{Joanne Gray}
\affiliation{%
  \institution{University of Sydney}
  \city{Sydney}
  \country{Australia}}
\email{joanne.gray@sydney.edu.au}

\author{Asangi Jayatilaka}
\affiliation{%
  \institution{RMIT University}
    \city{Melbourne}
  \country{Australia}}
\email{asangi.jayatilaka@rmit.edu.au}

\author{Louise La Sala}
\affiliation{%
  \institution{Orygen, Centre for Youth Mental Health, University of Melbourne}
    \city{Melbourne}
  \country{Australia}}
\email{louise.lasala@orygen.org.au}

\author{Nalin Arachchilage}
\affiliation{%
  \institution{RMIT University}
    \city{Melbourne}
  \country{Australia}}
\email{nalin.arachchilage@rmit.edu.au}

\author{Ryan M. Kelly}
\affiliation{%
  \institution{RMIT University}
    \city{Melbourne}
  \country{Australia}}
\email{ryan.kelly@rmit.edu.au}

\author{Sanchari Das}
\affiliation{%
  \institution{George Mason University}
  \city{Fairfax}
  \country{USA}}
\email{sdas35@gmu.edu}

\renewcommand{\shortauthors}{Wijenayake et al.}

\begin{abstract}
The growing prevalence of negative experiences in online spaces demands urgent attention from the human-computer interaction (HCI) community. However, research on online safety remains fragmented across different HCI subfields, with limited communication and collaboration between disciplines. This siloed approach risks creating ineffective responses, including design solutions that fail to meet the diverse needs of users, and policy efforts that overlook critical usability concerns. This workshop aims to foster interdisciplinary dialogue on online safety by bringing together researchers from within and beyond HCI --- including but not limited to Social Computing, Digital Design, Internet Policy, Cybersecurity, Ethics, and Social Sciences. By uniting researchers, policymakers, industry practitioners, and community advocates we aim to identify shared challenges in online safety research, highlight gaps in current knowledge, and establish common research priorities. The workshop will support the development of interdisciplinary research plans and establish collaborative environments --- both within and beyond Australia --- to action them.
\end{abstract}

\ccsdesc[500]{Human-centered computing}
\ccsdesc[300]{Human-centered computing~Interaction design}
\ccsdesc[300]{Human-centered computing~Collaborative and social computing}
\ccsdesc[300]{Social and professional topics~Computing / technology policy}
\ccsdesc[300]{Security and privacy~Human and societal aspects of security and privacy}

\keywords{online safety, digital harms, interdisciplinary collaboration, inclusive design, user-centred approaches, ethics, policy and governance, industry engagement}

\maketitle

\section{Introduction}

As online platforms increasingly mediate both our social lives and essential daily tasks — from connecting with others on social media to accessing government services — the issue of user safety has become more urgent than ever~\cite{esafety_commissioner_encounters_2025}. At the same time, negative experiences on online platforms are proliferating at an alarming rate. By 2021, 41\% of American adults reported having at least one negative online experience~\cite{vogels_state_2021}, whereas in Australia, the figure was significantly higher — with 70\% of adults experiencing at least one negative online interaction in the 12 months leading up to November 2022~\cite{esafety_commissioner_australians_2022}. Negative online experiences span a broad spectrum, including but not limited to hate speech and harassment~\cite{ping_perceiving_2025,agarwal_conversational_2024,ryu_exploring_2025}, impersonation~\cite{zhai_hear_2025,razaq_we_2021}, doxxing~\cite{goyal_you_2022,lee_use_2022,han_pressprotect_2024}, misinformation~\cite{malki_headline_2024,peng_leveraging_2024,wijenayake_effect_2020}, identity-based stereotyping~\cite{ardito_quantifying_2021,wijenayake_measuring_2019,alsawalqa_role_2022,felmlee_sexist_2020,wright_adolescents_2020}, non-consensual image sharing~\cite{hasan_your_2021,qiwei_sociotechnical_2024}, and financial scams~\cite{bellini_paying_2023,razaq_we_2021}. These experiences occur across a variety of digital services including social media~\cite{schulenberg_creepy_2023, schulenberg_creepy_2023,agha_strike_2023}, discussion forums~\cite{goyal_you_2022,lee_use_2022,han_pressprotect_2024}, messaging apps~\cite{agarwal_conversational_2024}, financial platforms~\cite{zhai_hear_2025,razaq_we_2021}, dating apps~\cite{aljasim_foregrounding_2023,datey_just_2024}, and online games~\cite{freeman_novel_2024,beres_dont_2021,madden_why_2021}. Moreover, the rapid adoption of generative AI has further intensified these challenges~\cite{freeman_new_2025,hawkins_deepfakes_2025}, introducing new threats such as deepfakes~\cite{umbach_non-consensual_2024, flynn_deepfakes_2022,henry_governing_2021} and large-scale misinformation campaigns~\cite{reuter_combating_2024,zhou_synthetic_2023} that not only extend existing problems but also create novel ways to target and deceive users. 

The current literature on online safety points to a clear conclusion: online safety is a multifaceted and increasingly complex problem that cannot be addressed through any single disciplinary lens. Yet research within HCI remains fragmented, with different sub-fields — including Social Computing, Digital Design, Internet Policy, Cybersecurity, Ethics, and Social Sciences — often tackling specific aspects of the problem in isolation~\cite{walker_what_2024}. Through this workshop, we aim to bring together researchers, policymakers, industry practitioners, and community advocates — predominantly within HCI and related disciplines — to critically examine how online safety is conceptualised and investigated across these domains. We will identify current approaches, methods, challenges, and knowledge gaps, and foster interdisciplinary collaboration aimed at advancing more comprehensive, inclusive, effective responses to online safety. A key goal of the workshop is to support the development of a collaborative community of online safety researchers, both within Australia and internationally.

\section{Rationale for the Workshop}

Online safety has emerged as a significant area of research across multiple sub-fields within HCI, each approaching the issue from different perspectives. For instance, research in CSCW and Social Computing has taken significant efforts to profile online risk experiences, understand users' safety perceptions, and develop targeted interventions to counter online harms~\cite{alsoubai_profiling_2024,obajemu_towards_2024,jean_baptiste_teens_2023,choe_mapping_2024}. Digital design and UX research has explored the usability and accessibility of safety tools, often highlighting barriers to adoption~\cite{agha_nudge_2023,agha_tricky_2024}. Participatory and co-design approaches have centred the experiences of users --- especially marginalised communities --- in the development of inclusive safety interventions~\cite{shanmugam_empowering_2024,agha_case_2022,sala_online_2025,aliyu_participatory_2024}.

Moreover, policy work has examined online safety at a systemic level, interrogating platform governance, content moderation practices, regulatory frameworks, and the protection of digital rights~\cite{trengove_critical_2022,phippen_online_2025,moran_end_2025}. In parallel, cybersecurity research has examined online safety by investigating how users perceive, interpret, and respond to digital threats such as phishing, with a focus on risk awareness, decision-making, and the design of effective training and educational interventions~\cite{sarker_multi-vocal_2024,beu_falling_2023,arachchilage_security_2014}.

Furthermore, Critical and Feminist HCI scholars have highlighted how structural inequalities such as gender, race, and ability shape both people's exposure to online harm and the effectiveness of safety interventions~\cite{matthews_supporting_2025,randazzo_trauma-informed_2023,jovic_ai_2025}. This literature focuses on the perspectives of marginalised users and questions whose needs are prioritised in the design of online platforms. Meanwhile, social sciences has shed light not only on the emotional and behavioural consequences of online harm --- such as trauma, coping, and bystander inaction --- but also on the psychological drivers of perpetration, victim blaming, and social dynamics that enable or discourage intervention~\cite{flynn_workplace_2024,flynn_intersectional_2024,flynn_attitudes_2025,felmlee_sexist_2020,wright_adolescents_2020}, all of which are key to designing responsive and effective safety interventions.

Despite this rich and growing body of work, current research efforts remain largely siloed. To move toward more collaborative and impactful approaches to online safety, this workshop has four key objectives, each of which is based on a limitation of the current research landscape.

\begin{description}
     
\item [Objective 1:] \textbf{Build a shared understanding of the current research landscape and promote mutual awareness across different disciplines.}
There is limited visibility across HCI sub-fields and other related disciplines regarding ongoing research in the online safety domain. Researchers tackling similar problems from different disciplinary perspectives are often unaware of parallel efforts, leading to duplicated work and missed opportunities for collaboration.

\item [Objective 2:] \textbf{Identify research gaps and collaborative opportunities.}
Research on online safety is frequently conducted in disciplinary silos, making it difficult to recognise shared challenges or determine where interdisciplinary collaboration could generate new insights. A more comprehensive view will help participants engage with broader problem spaces and discover new directions for impactful work.

\item [Objective 3:] \textbf{Initiate the development of an interdisciplinary research agenda that reflects the complexity and urgency of online safety issues.}
Although many individual projects address important facets of online safety, their scope, methods, and target audiences are often unaligned. A coordinated agenda can help connect these efforts, articulate shared priorities, and direct future research toward high-impact areas.

\item [Objective 4:] \textbf{Establish interdisciplinary networks to support ongoing, multi-stakeholder collaboration.}
Sustained collaboration between researchers, policymakers, industry practitioners, and community advocates remains rare, despite the complementary expertise each brings. Building these networks is crucial to translating research into real-world impact and implementing the proposed research agenda effectively.
\end{description}

\section{Research Areas of Interest}

Through structured discussions and collaborative activities, this workshop aims to promote conversation around interdisciplinary approaches to online safety, identifying how HCI research can address key challenges in the field. We invite participation from researchers, policymakers, industry practitioners, and community advocates working on online safety through the lenses of design, policy, technology, ethics, and the social sciences. Contributions are encouraged across the following key areas of online safety research (the full list of topics is provided in Section~\ref{topics}):

\begin{itemize}
    \item \textbf{Designing for Safety and Wellbeing:} User-centred and participatory design approaches to create inclusive, effective safety interventions.
    \item \textbf{Technology, AI, and Ethics:} Challenges and innovations in AI-driven safety tools, privacy, transparency, and ethical considerations.
    \item \textbf{Policy, Governance, and Industry:} Regulatory frameworks, cross-platform strategies, and industry perspectives on implementing safety measures.
    \item \textbf{Social, Behavioural, and Cultural Perspectives:} Understanding online harm through social dynamics, cultural contexts, and behavioural interventions.
\end{itemize}

We particularly encourage submissions that describe interdisciplinary work involving multiple stakeholders across academia, industry, policymakers, and community advocates to address the complex social, technical, policy, and ethical challenges of online safety. 

\section{Pre-workshop Plans}

\subsection{Workshop Website} 

The workshop website will be hosted at \url{www.talkingonlinesafety.org}. It will serve as the central hub for all workshop-related information. Prior to the event, we will publish the Call for Participation (see Section~\ref{cfp}) on the website, including the workshop objectives, topics of interest, submission instructions, key dates, and organiser profiles. The website will also provide a link to an Expression of Interest (EOI) form for interested participants to complete. The organisers will promote the Call for Participation through professional networks, university mailing lists, research groups, and social media. The organisers will also directly contact potential participants who may be interested, especially those from policy, industry, and community advocacy backgrounds, to ensure diverse representation. 

\subsection{Expression of Interest (EOI) Form} 
Those who intend to participate should complete an Expression of Interest (EOI) form (URL: \url{https://forms.gle/kYF3dwhhRndGENA9A}) by the submission deadline. The EOI form asks participants to provide their name, contact email, affiliation, research or professional area (or sub-field within HCI), accessibility requirements, a brief bio (maximum 300 words) outlining their relevant experience, and an abstract (maximum 300 words) describing their current or recent research, emerging problems, or new perspectives on online safety, as well as what they hope to gain from the workshop.

\subsection{Selection Process} The organising committee will review submissions and select participants based on diverse expertise, perspectives, and skills, as well as the quality, novelty, and relevance of their submissions. We aim to select participants to ensure diverse and inclusive representation across career stage, disciplinary background, and stakeholder group, including academia, industry, policy, and community advocacy. We plan to accept 15 - 20 workshop participants. Selected participants will be notified via email by the notification date. Once participants are confirmed, we will post a list of participants and their abstracts on the website to encourage early engagement and visibility. 

\section{Workshop Structure}

This full-day workshop (09:00 - 16:15) is structured in two sessions: a morning session focusing on small group discussions to identify and map participants' research backgrounds, interests, and challenges, and to explore opportunities for interdisciplinary, multi-stakeholder research into online safety. The afternoon session is dedicated to collaborative brainstorming of potential research proposals that participants can develop and continue exploring after the workshop.

\subsection{Inclusion and Accessibility Considerations}

The organising committee is committed to fostering an inclusive and accessible workshop environment. We will adhere to the conference’s accessibility and inclusion guidelines in all aspects of planning and delivery. A wheelchair-accessible venue has been requested, and participants will be invited to indicate any accessibility requirements in advance via the Expression of Interest form so that appropriate arrangements can be made. Workshop materials will be provided in accessible digital or printed formats, depending on participant preference. Activities are designed to support a range of participation styles — including verbal, written, and visual contributions — to accommodate neurodiverse participants and those with varied communication preferences.

\subsection{Morning Session}

\textbf{9:00 - 9:30 Workshop Opening \& Participant Profiles:} The session will begin with introductions from the workshop organisers, followed by an overview of the workshop objectives and agenda. We expect around 15–20 participants from fields such as HCI, Digital Design, Internet Policy, Cybersecurity, Ethics, and the Social Sciences, alongside policymakers, community advocates, and industry representatives. The workshop organisers are either from these disciplines or have established connections with these stakeholders and will leverage their networks to encourage participation, ensuring diverse perspectives.

Next, participants will create brief research profiles using a digital template provided by the organisers. These profiles will include their name, affiliation, contact information, research interests, methodological expertise, and any collaboration or funding opportunities they are exploring. The completed profiles will be accessible to participants during the workshop via a shared folder and will be distributed by email and posted on the workshop website after the event to support ongoing networking and collaboration.

\noindent\textbf{9:30 - 10:45 Mapping Participants' Research Interests:} Participants will be organised into interdisciplinary groups of 4–5 members, based on research interests identified through the pre-workshop EOI form. Each group member will present their research through a structured introduction covering their current work on online safety and its contribution to the field, the key research questions they are investigating, current challenges they face, and future opportunities they see for research. Participants may refer to their profiles during these presentations.

Following the introductions, the groups will engage in structured discussions using the following prompts:
\begin{itemize}
    \item How does each member's research address online safety concerns?
    \item What research questions are emerging from their different perspectives?
    \item What barriers are limiting progress in their areas?
    \item What unexplored opportunities exist for collaborative research?
\end{itemize}  

During these discussions, groups will collaboratively map their research landscape using sticky notes to visualise shared research questions, different approaches, and common challenges across disciplines.

\noindent\textbf{10:45 – 11:00 Morning Tea Break}

\noindent\textbf{11:00 – 12:00 Group Presentations:} Each group will present their discussion findings in 5-minute presentations. Collectively, these presentations will provide an overview of the current research landscape, highlighting different disciplinary approaches and common challenges. During the presentations, workshop organisers will facilitate discussions around similarities and differences between groups, questions about various approaches, and potential areas for collaboration. This process will help all participants understand the broader research context and set the foundation for collaborative idea development in the afternoon session.

\noindent\textbf{12:00 – 13:00 Lunch Break} 

\subsection{Afternoon Session}
    
\noindent\textbf{13:00 – 14:00 Developing a Mini-Project Proposal:} After lunch, participants will remain in their morning groups to develop a project proposal addressing a specific research gap or challenge identified during earlier discussions. Proposals may draw on insights from the group presentations or broader workshop conversations. Each proposal should outline the project aims, proposed methodology, expected outcomes, and how each member’s expertise will contribute. Groups will create a slide deck of up to 3 slides to summarise their proposal.

The goal of this activity is to develop feasible, interdisciplinary research projects that participants are genuinely interested in pursuing. These proposals will serve as a foundation for ongoing collaboration beyond the workshop. After the event, summaries of all proposals and discussion outcomes will be circulated to attendees via email and posted on the workshop website.

\noindent\textbf{14:00 - 14:45 Project Presentations:} Groups will present their research proposals in 10-minute presentations to share innovative project ideas and receive constructive feedback from all workshop participants. Participants will be encouraged to provide feedback, suggest improvements, and identify potential connections between different project proposals.

\noindent\textbf{14:45 - 15:00 Afternoon Tea Break}

\noindent\textbf{15:00 – 16:00 Expert Panel Discussion:} A panel of online safety experts will reflect on the day's discussions to provide external perspective on the research directions discussed throughout the workshop. The panel will include 4-5 members representing academics, policy experts, community advocates, and industry representatives, comprising both organising committee members and invited guests recruited through the committee's networks. A workshop organiser will moderate the panel discussion for 45 minutes, followed by a 15-minute question and answer session where workshop participants can seek advice and explore potential partnerships with the expert panel members.

\noindent\textbf{16:00 – 16:15 Closing and Next Steps:} The workshop will conclude with a summary of key outcomes and identified collaborations, along with information on follow-up activities. The aim is to ensure participants leave with clear pathways to continue the collaborations and research directions developed during the workshop.

\section{Post-workshop Publication Plan}
After the workshop, organisers will compile the researcher profiles and project proposals into a digital format and share them with attendees via email and post on the workshop website. This resource is expected to facilitate ongoing collaboration among participants and attract other researchers interested in online safety to explore partnership opportunities with workshop attendees.

\section{Call for Participation: Advancing Interdisciplinary Approaches to Online Safety Research}
\label{cfp}

\begin{itemize}
    \item Sydney, Australia
    \item Workshop date: 30 November 2025
    \item Website: \url{www.talkingonlinesafety.org}
    \item Contact: Senuri Wijenayake, RMIT University [senuri.wijenayake@rmit.edu.au]
\end{itemize}

\subsection{Important Dates}

\begin{itemize}
    \item Call for Participation: 19 September 2025 
    \item Submission Deadline: 19 October 2025 23:59 (AoE)
    \item Notification of Acceptance: 31 October 2025 23:59 (AoE)
    \item Workshop Dates: 30 November 2025 09:00 - 16:15, Sydney (Australia)
\end{itemize}

\subsection{Workshop Overview}

Online safety is a critical and complex challenge that affects users across diverse digital platforms. Despite substantial research efforts within Human-Computer Interaction (HCI) and related fields, work remains fragmented across disciplines such as Social Computing, Design, Policy, Cybersecurity, Ethics, and Social Sciences. This workshop aims to unite researchers, policymakers, industry practitioners, and community advocates to bridge these silos. Together, participants will share knowledge, identify gaps, and foster interdisciplinary collaboration to develop comprehensive, inclusive, and impactful solutions addressing the multifaceted nature of online safety.

\noindent The workshop will facilitate dialogue to:

\begin{itemize}
    \item Build a shared understanding of the current research landscape.
    \item Identify research gaps and opportunities for collaboration.
    \item Develop coordinated interdisciplinary research agendas.
    \item Establish sustained networks for multi-stakeholder collaboration.
\end{itemize}

\subsection{Topics of Interest}
\label{topics}

Topics of interest include, but are not limited to:

\begin{itemize}

  \item \textbf{Designing for Safety and Wellbeing}
  \begin{itemize}
    \item User-centred approaches to understanding and responding to online abuse
    \item \textit{Safety by design} principles and their practical implementation across platforms
    \item Designing for resilience and digital literacy in online safety
    \item Participatory design methods for developing safety interventions with affected communities
    \item Prevention-focused approaches to online safety
    \item Evaluation methods for assessing safety intervention effectiveness
    \item Psychological impacts of online harm and trauma-informed design
    \item Cultural and contextual considerations in online safety design, particularly for marginalised groups
  \end{itemize}

  \item \textbf{Technology, AI, and Ethics}
  \begin{itemize}
    \item Emerging online safety challenges from generative AI and synthetic media
    \item Use of AI for enhancing online safety and harm detection
    \item Transparency and explainability in safety-related AI systems
    \item Privacy-preserving safety interventions
    \item Ethical considerations in developing safety technologies, especially for marginalised or vulnerable users
    \item Evaluating the impact and unintended consequences of AI-driven safety tools
  \end{itemize}

  \item \textbf{Policy, Governance, and Industry}
  \begin{itemize}
    \item Policy frameworks and regulatory approaches to online safety
    \item Inclusion of equity-focused safety standards and obligations for protecting marginalised users
    \item Industry perspectives on implementing and scaling safety features
    \item Cross-platform safety considerations and interoperability
    \item Collaborative models involving academia, industry, policymakers, and communities
    
  \end{itemize}

  \item \textbf{Social, Behavioural, and Cultural Perspectives}
  \begin{itemize}
    \item Social and behavioural dynamics of online abuse (perpetrator, bystander, and survivor perspectives)
    \item Cross-cultural differences in perceptions of safety, harm, and acceptable behaviour online
    \item Structural, cultural, and socioeconomic factors influencing online safety for underrepresented communities
    \item Psychological impacts of sustained exposure to online harm and strategies for resilience
    \item Online safety education, awareness, and digital literacy interventions from a behavioural perspective
    \item Collective action, social movements, and advocacy around online safety and digital rights
  \end{itemize}
  
\end{itemize}

We particularly encourage submissions that describe interdisciplinary work involving multiple stakeholders across academia, industry, policymakers, and community advocates to address the complex social, technical, policy, and ethical challenges of online safety.

\subsection{Who Should Participate?}

We welcome submissions from researchers and practitioners across diverse disciplines such as HCI, Social Computing, Design, Policy, Cybersecurity, Ethics, and Social Sciences. Policymakers, industry professionals, and community advocates engaged in digital safety are encouraged to join. Both emerging and established researchers working on new or ongoing online safety projects will find this workshop valuable.

\subsection{How to Participate}

To participate, please submit an Expression of Interest (EOI) using this form: \url{https://forms.gle/kYF3dwhhRndGENA9A}. Your submission should include an abstract of up to 300 words describing your current or recent research, theoretical discussions, emerging challenges, or new perspectives related to online safety. Contributions from both early-stage and completed work are welcome. Selected participants will be invited to join a collaborative, interactive workshop focused on co-developing research agendas and building lasting interdisciplinary networks.

\subsection{Organisers}

\textbf{Senuri Wijenayake} is a Lecturer in the School of Computing Technologies at RMIT University. Her interdisciplinary research spans Human-Computer Interaction (HCI), Social Psychology, and Design to advance online safety for marginalised communities, including women and gender-diverse users. She examines the unique harms these communities experience, critiques the limitations of current technological responses, and uses participatory design methods to co-develop tailored safety interventions. Senuri was recently awarded a research grant from the Australian Communications Consumer Action Network (ACCAN) to develop design strategies and policy recommendations addressing technology-facilitated abuse on social media, in collaboration with WESNET, the eSafety Commissioner, and the Victorian Pride Centre.

\textbf{Joanne Gray} is a Lecturer in Digital Cultures in the Discipline of Media and Communications, Faculty of Arts and Social Sciences. She is an interdisciplinary academic with expertise in digital platform policy and governance. Her research seeks to understand how digital platforms, such as Google/Alphabet and Facebook/Meta, exercise private power and explore relevant policy options. Dr Gray is also the Commissioning Editor for the journal Policy \& Internet.
   
\textbf{Asangi Jayathilaka} is a Lecturer in Cybersecurity and Software Systems at RMIT University, Melbourne. She brings extensive experience in interdisciplinary research and participatory design, including co-design, specifically focused on empowering marginalised communities to shape their digital technology experiences. Her work deeply engages with diverse groups, including individuals with cognitive impairments, women, gender-diverse people, and culturally and linguistically diverse users.

\textbf{Louise La Sala} is a Research Fellow at Orygen, Centre for Youth Mental Health at the University of Melbourne. Her research is focused on youth self-harm and suicide prevention, with a specific interest in the impact of social media on the mental health and well-being of young people. Dr La Sala is a lead researcher on the \#chatsafe program of work which aims to provide young people, educators, and families with the tools to communicate safely online about self-harm and suicide. Her work investigates the complex relationship between social media and youth mental health, and she brings unique expertise in developing effective strategies to promote online safety and prevent self-harm and suicide among young people. Her work has informed online safety and suicide prevention policy at both a national and international level and she regularly works with popular social media platforms.

\textbf{Nalin Arachchilage} is an Associate Professor in the School of Computing Technologies at RMIT University. His research focuses on usable security and privacy, particularly at the intersection of computer security and human-computer interaction. He has developed innovative approaches to online safety, including a game design framework to educate users on protecting themselves from phishing attacks. Nalin's work bridges cybersecurity, HCI, and software engineering to create effective, user-centred security solutions.
    
\textbf{Ryan M. Kelly} is an Associate Professor in the School of Computing Technologies at RMIT University. He is interested in designing to enable safe experiences for older adults online, particularly in the areas of social interaction, digital health and finance. His recent research has focused on designing for safe video calling experiences between older adults and conversational volunteers. Ryan's disciplinary background is in Applied Psychology, HCI and CSCW. 
   
\textbf{Sanchari Das} is an Assistant Professor in the Department of Information Sciences and Technology at George Mason University, USA. Her research focuses on usable security and privacy, and applied AI, particularly in domains such as IoT, AR/VR/MR, healthcare, finance, education, and everyday digital interactions. She directs the CAPS (Center for AI, Privacy, and Security) Lab and co-directs the Secure Realities Lab, where she brings an interdisciplinary approach combining AI/ML, human-computer interaction, and cybersecurity to design resilient socio-technical systems. Prior to academia, she worked in industry as a Security and Software Engineer at American Express, Infosys, and HCL Technologies, and also served as a User Experience Consultant and Global Privacy Adviser for various organizations.

\begin{acks}
    This work was supported by the Australian Communications Consumer Action Network (ACCAN). The operation of ACCAN is made possible by funding provided by the Commonwealth of Australia under section 593 of the Telecommunications Act 1997. This funding is recovered from charges on telecommunications carriers.
\end{acks}

\bibliographystyle{ACM-Reference-Format}
\bibliography{references}

\end{document}